\documentclass[aps,prb, 
twocolumn,
groupedaddress]{revtex4-2}

\usepackage{dsfont}
\usepackage{graphicx}
\usepackage{dcolumn}
\usepackage{braket}
\usepackage{hyperref}
\usepackage{amsmath}
\usepackage{amssymb}
\usepackage{tikz}

\begin{document}


\title{\textbf{Extended Ashkin-Teller transition in two coupled frustrated Haldane chains} 
}%

\author{Bowy M. La Rivi\`ere}%
\email{Contact author: b.m.lariviere@tudelft.nl}
\affiliation{ Kavli Institute of Nanoscience, Delft University of Technology, Lorentzweg 1, 2628CJ Delft, The Netherlands }
\author{Natalia Chepiga}%
\affiliation{  Rudolf Peierls Centre for Theoretical Physics, University of Oxford, Clarendon Laboratory, Oxford OX1 3PU, United Kingdom}

\date{\today}

\begin{abstract}
We report an extremely rich ground state phase diagram of two spin-1 Haldane chains frustrated with a three-site exchange and coupled by the antiferromagnetic Heisenberg interaction on a zig-zag ladder.
A particular feature of the phase diagram is the extended quantum phase transition in the Ashkin-Teller universality class that separates the plaquette phase, which spontaneously breaks translation symmetry, and the uniform disordered phase. 
The former is connected to the Haldane phase, stabilized by large inter-chain coupling, via the topological Gaussian transition.
Upon decreasing the inter-chain interactions, this intermediate disorder phase vanishes, giving place to a dimerized phase separated from the plaquette phase on one side via a non-magnetic Ising transition and from the Haldane phase on the other side by a topological weak first-order transition.
Finally, in the limit of two decoupled chains, we recover a quantum critical point that corresponds to two copies of the Wess-Zumino-Witten $\mathrm{SU(2)}_2$ criticality with a total central charge $c=3$.
\end{abstract}


\maketitle

\section{Introduction}\label{sec: Introduction}
In recent decades antiferromagnetic Heisenberg spin chains with competing interactions have attracted a lot of attention due to the vast variety of  exotic quantum phases and phase transitions stabilized by frustration.
Critical spin-1/2 Heisenberg chains show spontaneous dimerization in the presence of next-nearest-neighbor interactions \cite{majumdarNextNearestNeighborInteractionLinear1969} when the ratio of the nearest-neighbor (NN) and next-nearest-neighbor (NNN) coupling exceeds $J_2/J_1 \approx 0.2411$ \cite{okamoto_fluid-dimer_1992}.
By contrast, the Heisenberg spin-1 chain realizes a topologically non-trivial phase with gapped excitations -- the Haldane phase \cite{HALDANE1983464}.
This phase is adiabatically connected to the Affleck-Kennedy-Lieb-Tasaki (AKLT) state \cite{affleck_rigorous_1987}, with valence bonds singlets (VBS) allocated at every nearest-neighbor bond. 
The topological nature of the phase results in emergent spin-1/2 edge states that are protected by symmetry \cite{TKennedy_1990,PhysRevLett.65.3181}. 
Some types of frustration, such as biquadratic interaction $J_b \left( \mathbf{S}_{i}\cdot\mathbf{S}_{i+1} \right)^2$ \cite{barber_spectrum_1989,klumper_new_1989,xian_exact_1993,lauchli_spin_2006} or the three-body interaction $J_3[\left(\mathbf{S}_{i-1}\cdot\mathbf{S}_{i}\right)\left(\mathbf{S}_{i}\cdot\mathbf{S}_{i+1}\right) + \mathrm{h.c}.]$ \cite{michaudAntiferromagneticSpinSChains2012}, are know to lead to spontaneous dimerization.
The  three-site interaction is particularly interesting in this respect as it generates an exact dimerized state at $J_3 =J_1/[4S(S+1)-2]$ -- a generalization of the Majumdar-Ghosh point for an arbitrary spin $S$ \cite{michaudAntiferromagneticSpinSChains2012}.
Haldane and Affleck have shown that for a set of integrable models with generic spin $S$ transitions between uniform and spontaneously dimerized phases can be effectively described by the Wess-Zumino-Witten (WZW) SU(2)$_{2S}$ critical theory \cite{affleckCriticalTheoryQuantum1987b}. 
It turns out, however, that for large spins $S\geq 5/2$, unless the system is fine-tuned, the transition renormalizes to a lower-level WZW critical theory \cite{PhysRevB.105.174402,Chepiga_spin_3_chain}.

There exists a fundamentally different class of quantum phase transitions in Heisenberg spin-$S$ chains that does not belong to the WZW family -- non-magnetic transitions, also known as valence-bond-solid transitions. 
These transitions can be effectively described by a re-arrangement of dimer covering of the lattice and are not associated with the appearance of deconfined spinons. 
The simplest of the kind -- the Ising transition -- has been reported in frustrated spin-1 chain in the presence of next-nearest-neighbor interaction. 
The latter is responsible for stabilizing the NNN-Haldane phase once $J_2/J_1\gtrsim 0.75$ \cite{kolezhukConnectivityTransitionFrustrated2002,kolezhukVariationalDensitymatrixRenormalizationgroup1997}.
This is a topologically trivial uniform phase that can be understood as a pair of intertwined Haldane chains \cite{kolezhukConnectivityTransitionFrustrated2002}. 
In the presence of negative biquadratic coupling $J_b$ or positive three-site interaction $J_3$ the system undergoes a non-magnetic Ising transition between the disordered NNN-Haldane phase and the dimerized phase with spontaneously broken $\mathbb{Z}_2$ symmetry \cite{chepigaDimerizationTransitionsSpin12016a,chepigaCommentFrustrationMulticriticality2016a}. 
From a field theory perspective, the emergence of Ising degrees of freedom can be intuitively understood through conformal embedding \cite{zamolodchikovNonlocalParafermionCurrents1985b}: all operators in the WZW SU(2)$_2$ critical theory can be represented as a product of a free boson and Ising fields. 
In vicinity of the WZW SU(2)$_2$ critical point the transition takes place simultaneously in both the boson and Ising sectors, but away from it the two degrees of freedom might split and result in an isolated Ising transition \cite{chepigaDimerizationTransitionsSpin12016a,chepigaCommentFrustrationMulticriticality2016a}. 
This has been further confirmed in a recent study of the $J_1-J_3$ spin-1 chain perturbed by a single-ion anisotropy, where the WZW SU(2)$_2$ point, upon reducing the SU(2) symmetry down to U(1) splits into a Gaussian and Ising transitions \cite{Pronk_deconfined}.
Similar physics has been observed in the anisotropic spin-1/2 ladders \cite{oginoContinuousPhaseTransition2021,oginoSymmetryprotectedTopologicalPhases2021,fontaineSymmetryTopologyDuality2024}.
Remarkably, the Ising transition has also recently been reported in spin-3 $J_1-J_2-J_3$ chain\cite{Chepiga_spin_3_chain}, generalizing the concept of non-magnetic transitions beyond the spin-1 case.

Ising criticality is one of the simplest examples of the minimal models in conformal field theory (CFT) \cite{difrancescoConformalFieldTheory1997}. 
Other prominent examples of this family of critical theories are the three-state Potts and Ashkin-Teller \cite{kohmotoHamiltonianStudiesAshkinTeller1981} universality classes that describe spontaneous breaking of $\mathbb{Z}_3$ and $\mathbb{Z}_4$ symmetries respectively.
A recent study on a family of quantum loop models \cite{Z4_transition_QLM} -- a simplified spin-1 model with the Hilbert space that by construction is limited  to the singlet-sector only -- has suggested that a non-magnetic Ashkin-Teller transition might appear between the NNN-Haldane phase and $\mathbb{Z}_4$ ordered plaquette phases.
However, among non-magnetic transitions only the Ising one has been observed in SU(2)-symmetric antiferromagnetic spin chains so far, while opportunities to realize more complex non-magnetic transitions remains unexplored. 

\begin{figure}[h!]
    \includegraphics{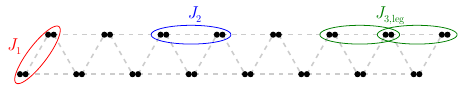}
    \caption{
       Sketch of a zig-zag ladder with interaction terms appearing in  Eq.\eqref{eq:Hamiltonian} marked: nearest-neighbor coupling $J_1$, next-nearest-neighbor coupling $J_2$, and the three body term $J_{3,\mathrm{leg}}$ acting on the legs of the ladder.
    }
    \label{fig: coupling}
\end{figure}

In this paper we address exactly this question and report the appearance of an extended Ashkin-Teller transition in two frustrated Haldane chains coupled in the zig-zag ladder. 
The model is defined by the following microscopic Hamiltonian:
\begin{equation}\label{eq:Hamiltonian}
    \begin{aligned}
        \mathcal{H} = \sum_i 
        & J_1 \mathbf{S}_{i}\cdot \mathbf{S}_{i+1} + J_2 \mathbf{S}_{i}\cdot \mathbf{S}_{i+2} \\
        & + J_{3,\mathrm{leg}} [\left(\mathbf{S}_{i-2}\cdot\mathbf{S}_{i}\right)\left(\mathbf{S}_{i}\cdot\mathbf{S}_{i+2}\right) + \mathrm{h.c}.],
    \end{aligned}
\end{equation}
where $J_1$ and $J_2$ are the typical NN and NNN antiferromagnetic Heisenberg interaction respectively, and $J_{3,\mathrm{leg}}$ is a three-site interaction acting on the legs as sketched in Fig.\ref{fig: coupling}. 
The latter appears along with next-nearest-neighbor and biquadratic interaction in the next-to-leading order expansion of the two-band Hubbard model \cite{michaudAntiferromagneticSpinSChains2012} and is responsible for exact dimerization.
The $J_1-J_2$ chain can be naturally viewed as a zig-zag ladder; in this picture the NN interaction corresponds to the inter-chain coupling and the three-site interaction defined in Eq.\ref{eq:Hamiltonian} acts as a frustration within each leg. 

When $J_{3,\mathrm{leg}}=0$ the system corresponds to a $J_1-J_2$ spin ladder realizing the topologically non-trivial Haldane phase when $J_1\gg J_2$ and the topologically trivial NNN-Haldane phase in the opposite limit,
separated by a first order transition at $J_1/J_2\approx1.3$ \cite{kolezhukVariationalDensitymatrixRenormalizationgroup1997,chepigaSpontaneousDimerizationCritical2016a,PhysRevB.101.115138}.
In the limit $J_1=0$ the system corresponds to a pair of uncoupled $J_1-J_3$ spin-1 chains previously studied in the literature \cite{michaudAntiferromagneticSpinSChains2012,chepigaDimerizationTransitionsSpin12016a,PhysRevB.88.224419}. 
When both $J_{3,\text{leg}}=0$ the system correspond to a pair of decoupled Haldane chains \cite{HALDANE1983464}, while at $J_{3,\mathrm{leg}}/J_2=1/6$ each chain is exactly dimerized. 
For a single chain, the transition between Haldane and dimerized phases takes place at $J_{3,\mathrm{leg}}/J_2\approx 0.111$ and belongs to the WZW SU(2)$_2$ universality class.
In the present case we have two copies of this critical theory resulting in a critical point with a particularly large central charge $c=3$ (later we will present the numerical results confirming this picture).

Inspired by the rich physics in the Majumdar-Ghosh spin-1/2 $J_1-J_2$ chain in the presence of inter-chain couplings, featuring, in particular, the non-magnetic Ising transition \cite{PhysRevB.95.144413}, we study the effect on inter-chain coupling on the generalization of the Majumdar-Ghosh chain to spin $S=1$.
We show that the Ashkin-Teller criticality appears in combination with the emergence of another topologically-trivial disordered phase in between the Haldane and $\mathbb{Z}_4$ ordered phase.
We study the ground-state phase diagram of this model, with open edges, numerically with a state-of-the-art two-site Density Matrix Renormalization Group (DMRG) algorithm \cite{whiteDensityMatrixFormulation1992a, whiteDensitymatrixAlgorithmsQuantum1993a}.
Without loss of generality we fix $J_2=1$ to set the energy scale.
We study chains of $N=4k$ sites, with $k$ denoting the number of plaquettes \footnote{Note that $N$ refers to the number of spin-1 sites and not the number of rungs of the zig-zag ladder.}.
Unless stated otherwise, we keep a maximum of $3000$ states, discard Schmidt values smaller than $10^{-8}$, and let the algorithm run till the absolute error in the energy of subsequent sweeps is smaller than $10^{-6}$.

The rest of the paper is structured as follows.
We start with  an overview of the phase diagram in Sec.\ref{sec: phase diagram}.
Then, in Sec.\ref{sec: Gaussian}, we discuss the transition between the intermediate disordered phase and the Haldane phase.
After that, we provide numerical evidence for the Ashkin-Teller transition in Sec.\ref{sec: Ashkin-Teller}.
In Sec.\ref{sec: Ising} we report the appearance of an Ising transition out of the $\mathbb{Z}_4$
Finally, we summarize our results and put them in a perspective in Sec.\ref{sec: Discussion}.


\section{Overiew of the phase diagram}\label{sec: phase diagram}
\begin{figure}[h!]
    \includegraphics{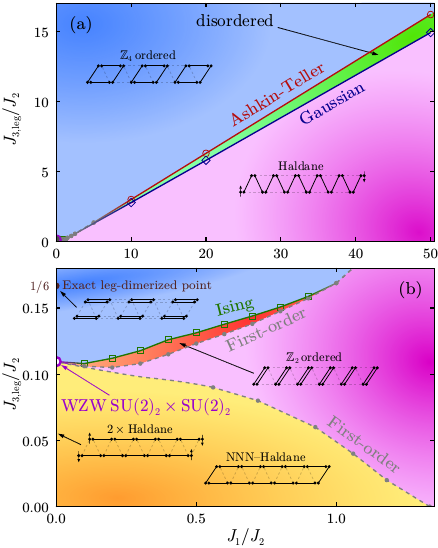}
    \caption{
        (a) Phase diagram of the spin-1 $J_1-J_2-J_{3,\mathrm{leg}}$ chain defined in Eq. \eqref{eq:Hamiltonian}. Each phase has a simple valence-bond-solid (VBS) representation sketched on the diagram. 
        A topologically trivial disordered phase (green) is separated on one side from the $\mathbb{Z}_4$ ordered phase (blue) by an Ashkin-Teller transition (red circles),
        and on the other side from the topological Haldane phase (magenta) by a Gaussian transition. Both transitions are characterized by the central charge $c=1$.
        The Haldane phase vanishes for $J_1 \lesssim 2$, and the transition between the $\mathbb{Z}_4$ ordered and Haldane phase becomes first order (grey circles) up to the point where intermediate $\mathbb{Z}_2$ ordered dimerized phase (red) appears. This is clearly visible in panel (b) that zooms in on $J_1$ small.
       The $\mathbb{Z}_2$ and $\mathbb{Z}_4$ ordered phases are separated by a continuous Ising transition. Haldane phase is separated from the dimerized and NNN-Haldane phases by a pair of first order transitions. At $J_1=0$ these transitions merge together with the Ising critical line into a multi-critical point that corresponds to two copies of WZW SU(2)$_2$ critical theory, separating the leg-dimerized phase from two uncoupled Haldane chains.
    }
    \label{fig: Phase diagram}
\end{figure}

In Fig.\ref{fig: Phase diagram} we present the phase diagram of the $J_1-J_2-J_{3,\mathrm{leg}}$ model defined in Eq.\eqref{eq:Hamiltonian}. We visualize each phase with the corresponding valence bond singlet (VBS) pattern covering the zig-zag lattice.
The phase diagram is extremely rich and contains multiple gapped.
We first consider the phases realized in the regime of strong NN and three-body interactions, as shown in Fig.~\ref{fig: Phase diagram}(a), of which there are three:
\begin{itemize}
    \item Topologically non-trivial Haldane phase. 
    In the limit $J_1 \gg J_2$ valence bond singlets are uniformly  distributed over the $J_1$ bonds (rungs) of the ladder. 
    Emergent spin-1/2 edge states are symmetry protected.
    \item $\mathbb{Z}_4$ ordered phase with a spontaneously broken translation symmetry. 
    In Fig.\ref{fig: Sketches}(a) we present a pattern of numerically extracted local correlations on the legs and rungs that supports this interpretation.  
    The phase includes a region dominated by leg dimerization, being exact at $J_1=0$ and $J_{3,\mathrm{leg}} = J_2/6 $ \cite{michaudAntiferromagneticSpinSChains2012}, and a region of plaquette order with a clear dimerized pattern along both legs and rungs. 
    The two regions -- leg dimerized and plaquette are described by the same spontaneously broken symmetry and are adiabatically connected, forming a single gapped phase.
    We show a sketch and a typical profile of the rung and leg correlations of the plaquette phase in Fig.\ref{fig: Sketches}(a).
    \begin{figure}[!b]
    \includegraphics{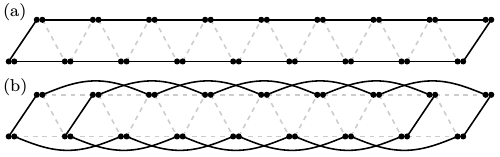}
    \caption{
        Illustrative VBS sketches of two possibilities of topologically trivial disordered phase that separates the Haldane and $\mathbb{Z}_4$ ordered phases.
        (a) The NNN-Haldane phase,
        (b) Fourth-neighbor-Haldane phase.
    }
    \label{fig: Uniform phase}
\end{figure}
    \begin{figure*}[t!]
    \includegraphics{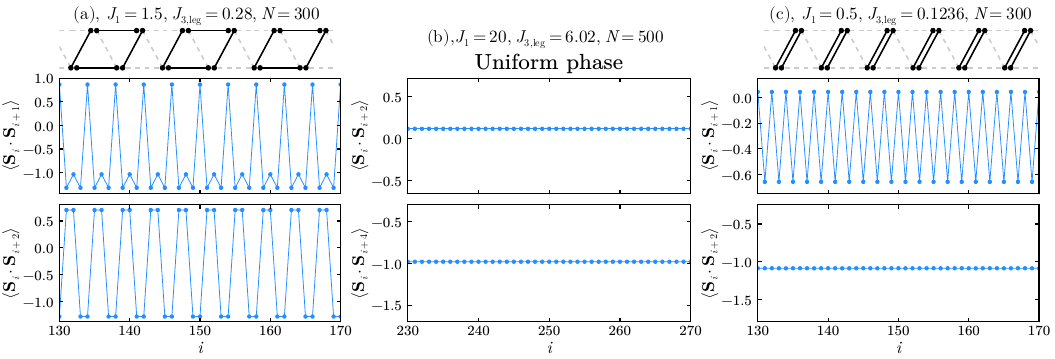}
    \caption{
        Typical local correlations for three different phases. 
        We also provide a sketch for the two ordered phases in a VBS format.
        (a) $\mathbb{Z}_4$ ordered plaquette phase.
        Broken bond order manifests itself in both rung $\langle \mathbf{S}_{i} \mathbf{S}_{i+1} \rangle$ and leg $\langle \mathbf{S}_{i} \mathbf{S}_{i+2} \rangle$ correlations. 
        (b) Intermediate topologically trivial disordered phase that separates the Haldane from the $\mathbb{Z}_4$ ordered phase.
        Leg correlations are weak while $\langle \mathbf{S}_{i} \mathbf{S}_{i+4} \rangle$ is strong.
        Both correlations are clearly uniform.
        A sketch of the two possibilities for this phase is shown in Fig.\ref{fig: Uniform phase}.
        (c) $\mathbb{Z}_2$ ordered dimerized phase. 
        Alternating order is found in the rung correlations while leg correlations does not show any order.
    }
    \label{fig: Sketches}
\end{figure*}
    \item Uniform topologically-trivial disordered phase. 
    For large $J_1$ an intermediate phase appears between the Haldane and $\mathbb{Z}_4$ ordered phase that lacks local order (see for instance the correlations in Fig.\ref{fig: Sketches}(b)).  
    It shows no signatures of topological order either. 
    There are two natural possibilities for this phase, as sketched in Fig.\ref{fig: Uniform phase}: {\it (i)} re-appearance of the NNN-Haldane phase, {\it (ii)} a phase with VBS singlets connecting NNN sites on every leg (thus fourth neighbor in the single-chain view). 
    The term $J_{3,\mathrm{leg}}$ can be decomposed into $\sum_{\alpha,\beta} S^\alpha_{i-2} Q^{\alpha \beta}_i S^{\beta}_{i+2} + \frac{4}{3}\mathbf{S}_{i-2} \mathbf{S}_{i+2}$, with $\alpha, \beta \in \{x, y, z\}$.
    The latter favors valence bond formation between sites $(i,i+4)$ resulting in a phase that could be dobbed fourth-neighbor-Haldane phase.
    From a symmetry perspective the two scenarios cannot be distinguished and are likely both represented in a true quantum ground state, though the state outlined in Fig.\ref{fig: Uniform phase}(b) is likely to have a dominant contribution. 
    This conclusion is based on the numerically extracted local correlations presented in Fig.\ref{fig: Sketches}(b), featuring a surprisingly strong $\langle \mathbf{S}_{i} \mathbf{S}_{i+4} \rangle$ compared to the weak NNN correlations $\langle \mathbf{S}_{i} \mathbf{S}_{i+2} \rangle$.
\end{itemize}

For the case of smaller NN and three-body interactions, shown in Fig.~\ref{fig: Phase diagram}(b): the phase diagram exhibits two additional gapped phases:
\begin{itemize}
    \item The topologically trivial NNN-Haldane phase. 
    In the limit of two uncoupled Haldane chains at $J_1=0$ this phase corresponds to a uniform distribution of valence bond singlets on $J_2$ bonds (legs) of the ladder. 
    In the presence of inter-chain coupling $J_1>0$ the system corresponds to a pair of intertwined Haldane chains \cite{kolezhukConnectivityTransitionFrustrated2002}. The system has no symmetry protected edge states.
    \item Dimerized phase.
    For  $J_1 \lesssim 1$ the Haldane and plaquette phases are separated by an intermediate dimerized phase. This is supported by the strong alternating pattern in the local rung correlations presented in Fig.\ref{fig: Sketches}(c) with the leg correlations remaining uniform.
\end{itemize}


The phase diagram also contains a multitude of  quantum phase transitions:
\begin{itemize}
    \item An Ashkin-Teller transition with central charge $c=1$ that covers a remarkably extended interval of the boundary between the intermediate uniform disordered phase and the $\mathbb{Z}_4$ ordered plaquette phase. 
    \item Gaussian transition separating the disordered phase from the Haldane phase. This is topological transition also characterized by the central charge $c=1$.
\item When this intermediate phase vanishes, the transition between the Haldane and plaquette phase becomes direct.
There are not enough symmetry-allowed relevant operators that could fine-tune the system into a direct and continuous transition between the Haldane and the  $\mathbb{Z}_4$ ordered plaquette phase resulting in an interval of the first order transition separating these two phases (see Appendix \ref{Appendix: first order Haldane and plaquette}).
\item Approaching the $J_1=0$ line, an Ising transition with $c=1/2$ branches off from this first order line. This continuous non-magnetic transition separates the plaquette phase from the $\mathbb{Z}_2$ ordered dimerized phase.
\item There are two first-order transitions separating the Haldane phase from the dimerized  (see Appendix \ref{Appendix: first order Haldane and dimerized}) and NNN-Haldane phases \cite{kolezhukConnectivityTransitionFrustrated2002,kolezhukVariationalDensitymatrixRenormalizationgroup1997}.
\item In the limit of two uncoupled frustrated Haldane chains, this first order line, alongside the Ising transition and the first order line at the other boundary of the dimerized phase, fuse into a WZW $\mathrm{SU}(2)_2 \times \mathrm{SU}(2)_2$ point that is characterized by $c=3$.
This special point separates the two uncoupled Haldane chains from the pure leg-dimerized phase at $J_1/J_2 \approx 0.111$ \cite{michaudAntiferromagneticSpinSChains2012}.
\end{itemize}
Let us now discuss each of these transitions in more details. Before diving into the key finding of the paper -- the extended Ashkin-Teller transition between the $\mathbb{Z}_4$ ordered plaquette and the uniform disordered phases -- we will first discuss the topological Gaussian transition as an important milestone in identifying the emergence of this uniform disordered phase in the first place.


\begin{figure*}[t!]
    \includegraphics{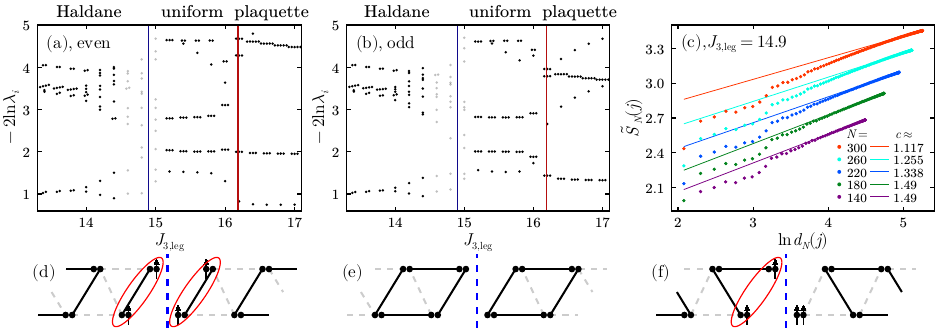}
    \caption{
        Evidence for a Gaussian transition between the topological Haldane phase and the topologically trivial uniform phase.
        (a) Entanglement spectrum as a function of $J_{3,\mathrm{leg}}$ for a chain of $N=200$ sites with open boundaries.
        We show the 10 lowest levels across the Haldane, uniform and the $\mathbb{Z}_4$ ordered phase. The data points (black dots) are slightly shifted horizontally so the degeneracy of the levels are clearly visible.
        Red and blue lines indicate the estimated location of the Gaussian and Ashkin-Teller transitions correspondingly. 
        Bi-partition is made (a) across  an even bond between sites $N/2$ and $N/2+1$ and (b) across an odd bond between sites $N/2+1$ and $N/2+2$.
        Data in (a)-(c) are shown for $J_1=50$.
        Note that the multiplicity of the spectrum in the uniform phases is unaltered, in contrast to those for the $\mathbb{Z}_4$ ordered phase.
        (c) Scaling of the reduced entanglement entropy with the logarithm of the conformal distance $d_N(j)$.  Colored dots denote different chain lengths.
        Data of all chain lengths but $N=140$ are shifted vertically for visual clarity.
        The central charge $c$ is extracted by fitting the data to Eq.\ref{eq:entanglement_entropy}; its numerical value approaches $c=1$ of the Gaussian transition upon approaching the thermodynamic limit.
        (d)-(f) Sketches of bi-parition of the $\mathbb{Z}_4$ ordered phase across (d)-(e) even and (f) odd bonds resulting in (d)-(e) unique and (f) three-fold degenerate lowest state in the entanglement spectrum.
    }
    \label{fig: Gaussian}
\end{figure*}

\section{Gaussian transition and the uniform disordered phase}\label{sec: Gaussian}
We start our analysis with the Gaussian transition that separates the topologically non-trivial Haldane phase from the disordered uniform phase that, as we will argue, lacks topological order.
These phases cannot be distinguished by any local order parameter.
In principle, the non-local string order parameter captures the destruction of the topological properties of the Haldane phase \cite{dennijsPrerougheningTransitionsCrystal1989,kennedyHidden$mathrmZ_2$ifmmodetimeselsetexttimesfi$mathrmZ_2$Symmetry1992,Kolezhuk_1996}, but its evaluation is numerically challenging. 
However, this difference in topological properties manifests itself in the entanglement spectrum as well \cite{Pollman_2010_spectrum}.
To compute this spectrum, we start with partitioning a quantum state $| \psi \rangle$ into two blocks with a Schmidt decomposition:
\begin{equation}\label{Schmidt}
    | \psi \rangle = \sum_i \lambda_i | L_i \rangle | R_i \rangle,
\end{equation}  
where $|L_i\rangle $ and $|R_i \rangle$ are the orthonormal basis of the left and right partition. 
The Schmidt values $\lambda_i$ can be calculated by tracing out one of the partitions and diagonalizing the remaining reduced density matrix $\rho_j$ \cite{li_entanglement_2008,levin_detecting_2006,kitaev_topological_2006}.
The multiplicity of these Schmidt values reflects the number of 
imaginary edge states that are created at the bi-partition. 
For topologically non-trivial phases with spin-1/2 edge states, such as the Haldane phase, all levels of the spectrum are always two-fold (strictly speaking, even-fold) degenerate \cite{Pollman_2010_spectrum} while for disordered phases without any topological order the levels in the entanglement spectrum might appear as non-degenerate or with an odd degeneracies \cite{chepigaSpontaneousDimerizationCritical2016a}.

In Fig.\ref{fig: Gaussian}(a) and (b) we show the entanglement spectrum $-2 \ln \lambda_i$ extracted from the ten largest Schmidt values $\lambda_i$ with $1 \leq i \leq 10$ for a bi-partitions made across an even and odd bond respectively.
The uniform phase clearly has a unique ground state for both the even and odd bi-partitions, confirming the lack of topological properties.
On the other hand, in agreement with previous works, the Haldane phase has a two-fold degeneracy across the entire spectrum, when far from the Gaussian transition.
Upon approaching this transition,  the degenerate levels are splitting due to the diverging correlation length that becomes of the same order as the length of the chain (see Appendix \ref{Appendix: Gaussian}), effectively coupling the spin-1/2 degrees of freedom localized at the edges with the effective fractional degrees of freedom caused by a bi-partition.

Let us also stress that the uniform phase has a non-degenerate spectrum for all bi-partitions, in contrast to the topologically-trivial $\mathbb{Z}_4$ phase that, although being topologically trivial, breaks translation symmetry and features an alternating non-degenerate and three-fold degenerate entanglement depending on the location of the bi-partition cut. 
Bi-partitioning across bonds such that there are an even number of sites on the left and goes through a plaquette, as sketched in Fig.\ref{fig: Gaussian}(d,) creates two spin-1/2 degrees of freedom that couple antiferromagnetically, resulting in a unique state. 
Bi-partitioning between the plaquettes, as sketched in Fig.\ref{fig: Gaussian}(e), also naturally leads to a non-degenerate entanglement spectrum. 
However, if we make the bi-parition across odd bonds, cutting one angle of a plaquette as sketched in Fig.\ref{fig: Gaussian}(f) at least on one side of the bi-partition, there is an unpaired spin-1 degrees of freedom resulting in a three-fold degenerate low-lying level of the entanglement spectrum.
The results presented in Fig.\ref{fig: Gaussian} (a)-(b) are fully compliant with this intuitive picture.

To confirm the Gaussian nature of the transition we first locate the critical point by looking at the peak in the divergent correlation length $\xi$ (see Appendix \ref{Appendix: Gaussian} for details) and the entanglement spectrum, and then we extract the central charge $c$ at the given critical point. 
For the latter, we extract the entanglement entropy $S_N(j)$ from the reduced density matrix $\rho_j$ as $ S_N(j) = - \mathrm{Tr} \rho_j \ln \rho_j $.
Then we calculate the reduced entanglement entropy by removing Friedel oscillations caused by the open ends of the chain \cite{laflorencie_boundary_2006,capponi_quantum_2013}:
$\tilde{S}_N(j) = S(j) - \zeta \langle \hat{S}^z_{j} \hat{S}^z_{j+1} \rangle$, where $\zeta$ is a non-universal constant. 
Finally, we fit $\tilde{S}_N(j)$ to the Calabrese-Cardy formula \cite{calabreseEntanglementEntropyQuantum2004}:
\begin{equation}
    \label{eq:entanglement_entropy}
    \tilde{S}_N(j) = \frac{c}{6} \ln d_N(j) + s_1 + \ln(g)
\end{equation}
where the pre-factor $c$ is the central charge, $d_N(j) =\frac{2N}{\pi}\sin\left( \frac{\pi j}{N}\right)$ is the conformal distance, $\ln g$ is a boundary entropy and $s_1$ is a non-universal constant.
We present a typical example of $\tilde{S}_N(j)$ as a function of $\ln d_N(j)$ in Fig.\ref{fig: Gaussian}(c) for a range of different chain lengths.
By fitting the slopes of the scaling -- here and in the rest of the article we always discard $30\%$ of sites near both edges -- we numerically extract central charges that are in good agreement with $c=1$ of the Gaussian universality class.


\begin{figure*}
    \includegraphics[width=0.8\textwidth]{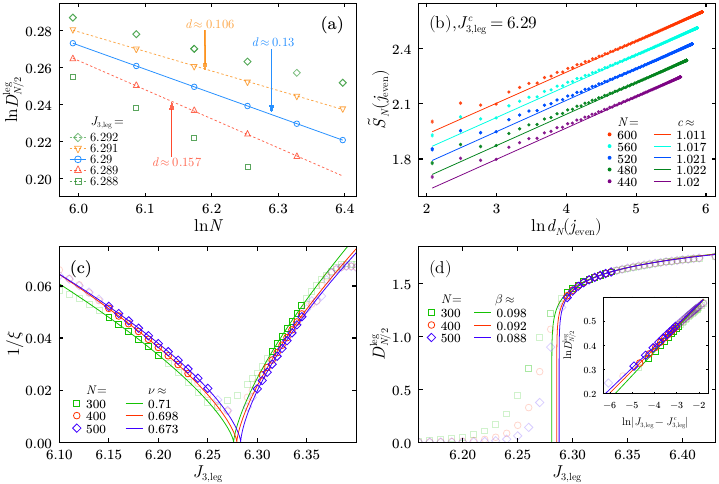}
    \caption{
        Numerical evidence for an Ashkin-Teller transition between the uniform phase and the $\mathbb{Z}_4$ ordered phase.
        Presented example is for $J_1=20$.
        Symbols show DMRG data, 
        lines state for the fits.
        (a) Finite-size scaling of the mid-chain leg dimerization as defined in Eq.\eqref{eq:dimerization_order_parameter} in a log-log plot.
        We associate a critical point with a separatrix at $J_{3,\mathrm{leg}}^c \approx 6.29$;  its slope corresponds to a scaling dimension $d \approx 0.130$ that is in excellent agreement with the universal $d=1/8$ of the Ashkin-Teller critical theory.
        Concave and convex curves are shifted vertically for visual clarity.
        (b) Scaling of the reduced entanglement entropy with conformal distance at the critical point $J_{3,\mathrm{leg}}^c$ identified in (a).
        We only show data (dots) for bi-partitions made across even bonds.
        Curves for $N>440$ are shifted for visual clarity. 
        By fitting the data with the Eq.\ref{eq:entanglement_entropy} we extract the central charge $c$ that is in excellent agreement with $c=1$. 
        (c) Inverse of the correlation length for three system sizes.
        Lines mark the fit with $\xi \propto |J_{3,\mathrm{leg}}-J_{3,\mathrm{leg}}^c|^{-\nu}$; numerically extracted values of $\nu$ fits well the  interval $2/3 \leq \nu \leq 1$ of Ashkin-Teller critical theory.
        Data with $\xi > N/10$, or those far away from the critical point, are excluded from the fit (pale symbols).
        (d) Mid-chain dimerization on the legs near the critical point.
        Curves show a fit with $D^\mathrm{leg}_{N/2} \propto (J_{3,\mathrm{leg}}-J_{3,\mathrm{leg}}^c)^{\beta}$. Numerically extracted critical exponent $\beta$ matches well the Ashkin-Teller interval $1/12 \leq \beta \leq 1/8$. Inset: Same data but in a log-log scale.
           }
    \label{fig: Ashkin-Teller}
\end{figure*}

\section{Extended Ashkin-Teller transition}\label{sec: Ashkin-Teller}

Let us now focus on the transition at the other side of the topologically-trivial uniform phase that separates it from the $\mathbb{Z}_4$ ordered plaquette phase.
Spontaneous breaking of $\mathbb{Z}_4$ symmetry is typically associated with the Ashkin-Teller universality class \cite{kohmotoHamiltonianStudiesAshkinTeller1981,difrancescoConformalFieldTheory1997} -- a family of critical theories with continuously varying critical exponents.
This {\it weak} universality class is characterized by a universal central charge $c=1$ and scaling dimension $d=\beta/\nu=1/8$ of the order parameter associated with the symmetry broken phase.   
However, other critical exponents, such as $\nu$ and $\beta$ are not unique (though their ratio is) and might change along the transition. 
Their values interpolate from a pair of decoupled Ising chains with $\nu=1$ and $\beta=1/8$ to the symmetric 4-state Potts point with $\nu=2/3$ and $\beta=1/12$ \cite{kohmotoHamiltonianStudiesAshkinTeller1981,aounPhaseDiagramAshkin2024}. 
In the remainder of this section we will compute all four of these quantities to confirm the nature of the transition between the uniform and the plaquette phases.

We start with the universal scaling dimension $d$ as it also provides us a way for accurate estimate of the location of the transition. 
We associate the order parameter with leg correlations  $\langle \mathbf{S}_i \cdot \mathbf{S}_{i+2} \rangle$: it  shows a clear four-site alternating pattern in the plaquette phase while being uniform in the intermediate disordered phase (for typical examples see the bottom row of Fig.\ref{fig: Sketches}(a) and (b)). 
We define the corresponding leg dimerization order parameter as:
\begin{equation}\label{eq:dimerization_order_parameter}
    D^\mathrm{leg}_i = \left|\left\langle \mathbf{S}_{i} \cdot \mathbf{S}_{i+2} - \mathbf{S}_{i+2} \cdot \mathbf{S}_{i+4} \right\rangle \right|
\end{equation}
that we compute in the middle of the chain $i=N/2$ to minimize boundary effects.
Since this operator is symmetric with respect to the top or bottom chain in the zig-zag ladder, from now on we will present it for the top chain only.
In the thermodynamic limit $D^\mathrm{leg}_{N/2}$ assumes a finite value in the plaquette phase, and scales to zero in the disordered phase.
As a function of the length of the chain $N$, these respectively appear as concave and convex curves in a log-log plot. 
The separatrix in between these curves marks the critical point, and according to boundary CFT the order parameter $D^\mathrm{leg}$ scales algebraically with the length of the chain $D^\mathrm{leg}_{N/2} \propto N^{-d}$ \cite{difrancescoConformalFieldTheory1997}. 
We show a typical example of such a finite size scaling for $J_1=20$ in Fig.\ref{fig: Ashkin-Teller}(a).
By fitting the slope of the separatrix we extract a scaling dimension $d \approx 0.130$, which agrees within $4\%$ interval with the theoretical value $d=1/8$ of the Ashkin-Teller universality class.

At the critical point we computed the reduced entanglement entropy $\tilde{S}_N(j)$. It scales linearly with the logarithm of the conformal distance $d_N(j)$ as shown in Fig.\ref{fig: Ashkin-Teller}(b).
Note, we only show data for bi-partitions occurring across even bonds (where we cut either through the center of a plaquette or right in between two of them \footnote{$\langle \mathbf{S}_{i} \cdot \mathbf{S}_{i+2} \rangle $ correctly removes Friedel oscillations in this case while bi-partitioning the other bonds requires a combination of both $\mathbf{S}_{i} \cdot \mathbf{S}_{i+1}$ and $\mathbf{S}_{i} \cdot \mathbf{S}_{i+2}$.}).
By fitting the data to the Calabrese-Cardy formula (Eq.\eqref{eq:entanglement_entropy}) we extract a central charge $c$ that falls within a $3\%$ error margin of $c=1$ of the Ashkin-Teller critical theory.

The critical exponents $\nu$ and $\beta$ respectively describe the divergence of the correlation length $\xi \propto |J_{3,\mathrm{leg}}-J_{3,\mathrm{leg}}^c|^{-\nu}$ and the scaling of the order parameter $D^\mathrm{leg}_{N/2} \propto |J_{3,\mathrm{leg}} - J_{3,\mathrm{leg}}^c|^{\beta}$ upon approaching the transition.
To calculate $\nu$, we first extract $\xi$ by computing the $z$-component of the connected leg-leg correlation function 
\begin{equation}\label{eq: connect correlations}
    C_{i,j} = \langle S^z_{i} S^z_{i+2} S^z_{j} S^z_{j+2} \rangle - 
            \langle S^z_{i} S^z_{i+2} \rangle \langle S^z_{j} S^z_{j+2} \rangle,
\end{equation}
and fitting it to:
\begin{equation}\label{eq: correlation length}
    C_{i,j} \propto \frac{e^{-|i-j|/\xi}}{\sqrt{|i-j|}}, 
\end{equation}
for multiple points close to the transition (see Appendix \ref{Appendix: correlation function} for more information).
We then fit the collected data points for $1/\xi$ as a function of $J_{3,\mathrm{leg}}$, as shown in Fig.\ref{fig: Ashkin-Teller}(c), to extract the numerical estimate of $\nu$ for a given system size.

To extract the order parameter critical exponent $\beta$ on the other hand, we fit $D^\mathrm{leg}_{N/2}$ in the $\mathbb{Z}_4$ ordered phase, treating the location of the critical point as a fitting parameter.
We show typical examples of both of this in Fig.\ref{fig: Ashkin-Teller}(d).
In addition, we also depict $D^\mathrm{leg}_{N/2}$ in the plaquette phase in vicinity to the critical point in a log-log scale (see inset), and find the expected scaling.
Both $\beta$ and $\nu$ agree well with $2/3 \leq \nu \leq 1$ (or equivalently $1/12 \leq \beta \leq 1/8$) that is characteristic of an Ashkin-Teller transition.
As an additional check, we obtain a scaling dimension $0.138 \lesssim d \lesssim 0.131$ from these $\nu$ and $\beta$, agreeing well with $d$ extracted from the finite size scaling of $D^\mathrm{leg}_{N/2}$.

\begin{figure}[t]
    \includegraphics{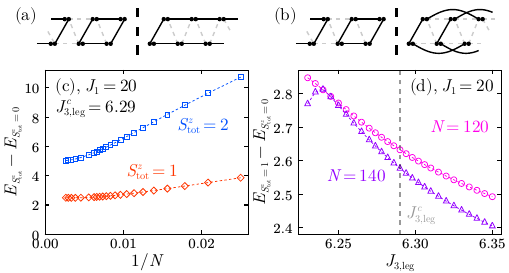}
    \caption{
        Proof of the non-magnetic nature of the Ashkin-Teller transition.
        (a) - (b) Sketches of the domain walls between the plaquette phase and the uniform disordered phase.
        We depict them for both possibilities of the disordered phase that are illustrated in Fig.\ref{fig: Uniform phase}.
        (c) Finite size scaling of the singlet-triplet gap $E_{S^z_\mathrm{tot}=1} - E_{S^z_\mathrm{tot}=0}$ (red diamonds) and the singlet-quintuplet $E_{S^z_\mathrm{tot}=2} - E_{S^z_\mathrm{tot}=0}$ gap (blue squares) at the critical point $J_{3,\mathrm{leg}}^c=6.29$ for $J_1=20$ in the thermodynamic limit (see finite size scaling of the order parameter in Fig.\ref{fig: Ashkin-Teller}).
        Both energy gaps show no tendency to close.
        (d) Singlet-triplet gap as a function of $J_{3,\mathrm{leg}}$ for two different chain lengths.
        Vertical dashed purple line shows the location of the critical point.
        Lack of a dip in the energy at the transition indicates a finite gap.
    }
    \label{fig: energy gap}
\end{figure}

Non-magnetic transitions are associated with the condensation of non-magnetic domain walls, i.e. those that do not carry spinors. 
Notable examples are the previously mentioned Ising transitions between topologically trivial uniform phases and spontaneously dimerized phases in spin-1 \cite{chepigaDimerizationTransitionsSpin12016a,chepigaCommentFrustrationMulticriticality2016a,chepigaDimerizationEffectiveDecoupling2019} and spin-3 chains \cite{Chepiga_spin_3_chain}.
Similar non-magnetic domain walls are realized at the interface between the uniform disordered and plaquette phases as sketched in Fig.\ref{fig: energy gap}(a) and (b).
At magnetic transitions the condensation of magnetic domain walls is reflected in the closing of magnetic energy gaps.
By contrast, in the present case we see that singlet-triple and singlet-quintuplet gaps remain open across the Ashkin-Teller transition, further supporting its non-magnetic nature. 
We present a finite-size scaling of the singlet-triplet energy gap at the critical point in Fig.\ref{fig: energy gap}(c). 
In Fig.\ref{fig: energy gap}(d) we show how finite-size gaps vary across the transition.  
In order to extract the singlet-triplet (quintuplet) gap we compare the lowest energy states in the sectors with a total magnetization $S^z_\mathrm{tot}=0$ and $S^z_\mathrm{tot}=1$ ($S^z_\mathrm{tot}=0$ $S^z_\mathrm{tot}=2$).
Based on these results, we conclude that the magnetic excitation energy remains finite and shows no tendency of closing at or near the transition \footnote{The non-magnetic excitations and the corresponding singlet-singlet gap is expected to close at the transition. However, due to strong finite size effects, direct numerical verification of this expectation is currently unfeasible.}.
In addition, we present the results for the singlet-quintuplet energy gap. It stays finite at the critical point with the value almost twice as large as the singlet-triplet gap, indicating that two magnetic excitations in the quintuplet state are local and hardly interact with each other.

Finally, let us emphasize the remarkable extent of the Ashkin-Teller transition.
Even though we were not able to determine the precise location of the start and end points of the Ashkin-Teller line, we observe clear signatures of Ashkin-Teller criticality between $J_1=10$ and $J_1=50$ (see Appendix \ref{Appendix: Ashkin-Teller}) indicating a lower estimate of its actual extent. 


\section{Ising transition}\label{sec: Ising}

\begin{figure}[!t]
    \includegraphics{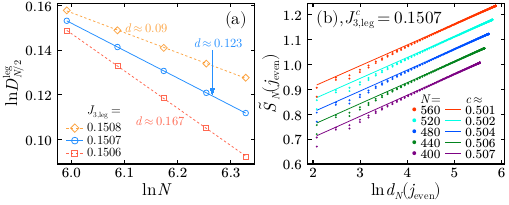}
    \caption{
        Numerical data for an Ising transition between the $\mathbb{Z}_2$ ordered dimerized phase and the $\mathbb{Z}_4$ ordered plaquette phase.
        Data is presented for $J_1=0.8$.
        (a) Finite size scaling of the dimerization on the legs, defined in Eq.\eqref{eq:dimerization_order_parameter}, in the middle of the chain.
        Separatrix denotes the critical point and scales with a scaling dimension $d \approx 0.123$ that is in excellent agreement with the theoretical value $d=1/8$.
        (b) Reduced entanglement entropy as a function of the logarithm of the conformal distance (colored dots).
        Data is shown for the critical point obtained in (a).
        We only show data for $j_{even}$: bi-partitions made through the middle of a plaquette, or between two of them.
        Curves are shifted for visual clarity except for the one corresponding to $N=400$.
        The extracted central charge agrees well with the CFT prediction $c=1/2$ for the Ising universality class.
    }
    \label{fig: Ising}
\end{figure}

Once the nearest-neighbor Heisenberg coupling is smaller than the next-nearest-neighbor one,  $J_1/J_2\lesssim 1$, we detect the appearance of another intermediate phase separating the $\mathbb{Z}_4$ plaquette and the Haldane phases (see Fig.\ref{fig: Phase diagram}(b)). 
By contrast to the uniform disordered phase that we have encountered for large $J_1$, the present phase breaks translation symmetry with a clear dimerization pattern on the rungs as shown in Fig.\ref{fig: Sketches}(a) and (c).
Similar to the previous case, the dimerized and the plaquette phases can be distinguished by the middle-chain leg dimerization $D_{N/2}^\mathrm{leg}$. We perform finite size scaling of this operator following the same procedure as before; these results are presented in Fig.\ref{fig: Ising}(a). From the slope of the separatrix we extract a scaling dimension $d\approx 0.123$ which agrees perfectly with the CFT prediction $d=1/8$ for the Ising criticality.

At the critical point we compute the reduced entanglement entropy, and its scaling with the conformal distance as shown in Fig.\ref{fig: Ising}(b).
The central charge that we extract from the scaling is in excellent agreement with $c=1/2$.
This confirms the Ising nature of the transition, and agrees with earlier predictions for the same transition in spin-1 quantum loop models \cite{Z4_transition_QLM}.

Intuitively the appearance of the Ising transition can be understood as one of the two-step process of getting into $\mathbb{Z}_4$ by spontaneously breaking $\mathbb{Z}_2$ twice. First, the symmetry between even and odd rungs is broken with dimers placed on either of the two in the dimerized phase. Second, the allocated rung dimers are grouped in pairs to form plaquettes and breaking the translation symmetry on legs. The latter corresponds to the Ising transition considered here. The former is the first order transition in the present case, but upon deforming the model it could be connected to a WZW SU(2)$_2$, that in turn can be understood as a coset of U(1) boson field (responsible for disappearance of the topological order) and $\mathbb{Z}_2$ Ising (associated with the appearance of dimers and translation symmetry broken by two sites) \cite{chepigaDimerizationTransitionsSpin12016a}.


\section{Discussion}\label{sec: Discussion}
To summarize, we report the appearance of a non-magnetic Ashkin-Teller transition in a system of two frustrated Haldane chains coupled in a zig-zag ladder.
We verified the universality class of this remarkably extended critical line against CFT predictions for the scaling dimension $d$ and the central charge $c$, as well as the non-universal critical exponents $\beta$ and $\nu$ that control the scaling of the order parameter and the divergence of the correlation length with the distance to the transition.
Our analysis is based on density matrix renormalization group calculations: the scaling dimension $d$ is obtained via finite-size scaling of the relevant operator, the central charge $c$ from the entanglement entropy, and the exponents $\beta$ and $\nu$ from the scaling behavior of the corresponding observables.

The realization of the Ashkin-Teller transition is intimately connected with the appearance of an intermediate disordered phase between the Haldane and plaquette phases. This phase, being topologically trivial and uniform, can be connected to the $\mathbb{Z}_4$ ordered domain through a spinor-free non-magnetic domain wall, resulting, in turn, into a non-magnetic Ashkin-Teller transition.
We confirmed this picture through a finite-size analysis of the low-lying magnetic spectra.
On the other side, however, the magnetic Gaussian transition separates the uniform intermediate phase from a topologically non-trivial Haldane phase. For smaller inter-chain coupling $J_1$ the two transitions approach each other, however, we do not see a direct $c=2$ transition where topological and $\mathbb{Z}_4$ critical lines would fuse into a single deconfined quantum critical line. Instead, we report a short interval of the first order transition separating the Haldane and the plaquette phases. It would be interesting, of course, to study if with additional interaction terms one could produce a direct continuous transition between these two phases, realizing a fine-tuned higher level WZW critical theory.
This exciting exploration, however, goes beyond the scope of the present work.

Speaking about additional terms, we expect a qualitatively similar phase diagram featuring, in particular a non-magnetic Ashkin-Teller transition, for two bilinear-biquadratic chains coupled in the same manner in a zig-zag ladder.
Negative biquadratic interaction is responsible for stabilizing the dimerized phase in a single chain \cite{lauchli_spin_2006}, and, as we have seen, leads to the emergence of the plaquette phase in the presence of inter-chain coupling. However, by contrast to the three-site interaction,  the biquadratic interaction does not lead to the exactly dimerized states \cite{lauchli_spin_2006,chepigaCommentFrustrationMulticriticality2016a,PhysRevB.96.134404}. The spontaneous dimerization for this model is overall significantly smaller, that makes its numerical investigation even more challenging.
Furthermore, the biquadratic interactions have to be comparable or exceeding the Heisenberg interactions within legs and thus cannot be classified as a perturbation.

\begin{figure}[!b]
    \includegraphics{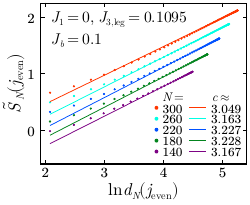}
    \caption{
        Proof for a WZW $\mathrm{SU}(2)_2 \times \mathrm{SU}(2)_2$ critical point between two uncoupled Haldane chains and the leg-dimerized phase with a broken $\mathbb{Z}_2 \times \mathbb{Z}_2$ order.
        We show the reduced entanglement entropy as a function of the logarithm of the conformal distance at the critical point $J_3/J_1 \approx 0.11$.
        We only show data corresponding to bi-partitions through two strong bonds or none, and a small antiferromagnetic coupling $J_b=0.1$ acts first and last rung. 
        Colored dots depict five different chain lengths, and for $N>140$ data is shifted for visual clarity.
        The extracted central charge agrees well with the expected $c=3$.
    }
    \label{fig: 2 times WZW}
\end{figure}

Now, let us look closely at the limit of two decoupled chains ($J_1=0$) with a single multi-critical point separating two Haldane chains from the purely leg-dimerized state.
In the single $J_1-J_3$ chain, the WZW SU(2)$_2$ point characterized by a central charge $c=3/2$ separates the dimerized phase from the Haldane chain at $J_3/J_1 \approx 0.11$ \cite{michaudAntiferromagneticSpinSChains2012}. 
Now we have two non-interacting copies of such chains and we thus expect a WZW $\mathrm{SU(2)}_2 \times \mathrm{SU(2)}_2$ critical point with $c=3$.
This is confirmed by the reduced entanglement entropy, computed for $J_{3,\mathrm{leg}} = 0.1095$, which we show in Fig.\ref{fig: 2 times WZW}.
It is instructive to look at this point from a perspective of conformal embedding 
\cite{zamolodchikovNonlocalParafermionCurrents1985b}. 
As briefly outlined in the introduction, the WZW SU(2)$_2$ critical theory can be formulated as  a product of the U(1) boson and Ising fields $\mathrm{U(1)} \times \mathbb{Z}_2$, with central charges $c=1$ and $c=1/2$ correspondingly adding up to $c=3/2$ of the WZW SU(2)$_2$ theory.
In the present case, we start with two copies of the WZW SU(2)$_2$ critical theories at $J_1=0$, that in the presence of inter-chain coupling splits into three transitions: the Ising transitions with $c=1/2$, the first order transition between Haldane and NNN-Haldane phases, and another first order transition between the Haldane and the dimerized phases. Previous studies have shown that the latter, if continuous, belongs to the WZW SU(2)$_2$ universality class but can be turned into a first order transition upon changing a sign of the marginal operator.  In the present case our results suggest a weak first order transition with incommensurate short-range correlations on both sides of it. 
However, additional terms balancing the marginal operator can potentially turn this transition into a continuous $c=3/2$ critical line. The transition between the Haldane and NNN-Haldane phases is also a weak first order one but following the logic of Ref.\onlinecite{chepigaDimerizationTransitionsSpin12016a} we can associate it with the U(1) boson field. In this picture sketched in Fig.\ref{fig: 2 times WZW}, one of the two WZW SU(2)$_2$ critical line turns into first order, while another one in accordance with conformal embedding decomposes into the critical Ising line and a separate U(1) boson transition that, however, is not continuous, but weak first order. For larger values of $J_!$ the dimerized phase disappears and the Ising transition fuses with the first order transition (associated with the WZW SU(2)$_2$ line). If the outcome would be a continuous transition, it would be (as pointed previously) a $c=2$ criticality. However, in the present case we always detect a direct first order transition between the Haldane and the plaquette phase.

\begin{figure}[!t]
    \includegraphics{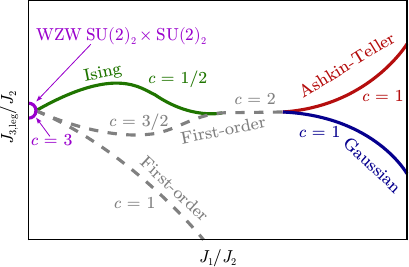}
    \caption{
    Schematic of the phase diagram shown in Fig.\ref{fig: Phase diagram} to emphasize the fusion of critical and first order lines.
    True central charges of continuous lines are depicted in their respective colors and the 'hidden' central charges of the first order lines are shown in gray. 
    }
    \label{fig: conformal embedding}
\end{figure}


Generalizing the concept of non-magnetic transitions beyond the simplest Ising case opens up promising avenues for further development. In contrast to the $\mathbb{Z}_2$ ordered phases, the melting of phases with higher periodicities — particularly the $\mathbb{Z}_3$ trimerized and $\mathbb{Z}_4$ plaquette phases — offers substantially richer critical behaviour. In addition to the conformal transitions, such as the Ashkin-Teller critical line reported here, there may be more exotic types of quantum criticality, including chiral transitions and non-magnetic floating phases, which are realized in response to chiral perturbations that are manifested through incommensurate correlations \cite{Huse_Fischer_1982,Huse_Fischer_1984, Chepiga_Hard_boson,chepiga_kibble-zurek_2021,Jose_Z4_chiral,Z4_transition_QLM}.  In the present case, we report the short-range incommensurate correlations throughout the uniform disordered phase (see Appendix \ref{Appendix: incommensurate oscillations}). However, incommensurability disappears shortly before the Ashkin-Teller transition keeping the chiral transitions in quantum magnets in the yet-to-be-observed status.

\begin{acknowledgments}
NC thanks to F.Mila for a collaboration on a related project on non-magnetic Ising transitions. This research was supported by Delft Technology Fellowship. 
The work of NC was funded by the European Union through the ERC grant (TRANGINEER, 101220181). Views and opinions expressed are however those of the author(s) only and do not necessarily reflect those of the European Union or the European Research Council Executive Agency. Neither the European Union nor the granting authority can be held responsible for them.
Numerical simulations were performed at the DelftBlue HPC and at
the Dutch national e-infrastructure with the support of the SURF Cooperative.
\end{acknowledgments}

\appendix

\section{First order transition between Haldane and plaquette phase}
\label{Appendix: first order Haldane and plaquette}

\begin{figure}[b]
    \includegraphics{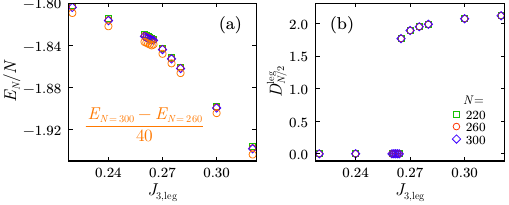}
    \caption{
        Numerical evidence for a first order transition between the topologically trivial $\mathbb{Z}_4$ ordered plaquette phase and the disorder Haldane phase with topologically protected edge states.
        Data is presented for $J_1=1.5$.
        (a) Ground state energy per site $E_N/N$ shows a clear kink for three different chain lengths.
        This kink is also apparent in an estimate of the bulk energy per site, that we compute as $(E_{N_1}-E_{N_2})/(N_1-N_2)$ (orange circles), and which has significantly reduced boundary effects.
        (b) The dimerization on the legs in the middle of the zig-zag ladder shows a clear jump.
    }
    \label{fig: first order Haldane and plaquette}
\end{figure}

The transition between the topologically non-trivial Haldane phase and the topologically trivial plaquette phase with a broken $\mathbb{Z}_4$ order shows clear signatures of a first order transition, as presented in Fig.\ref{fig: first order Haldane and plaquette}.
First, the ground state energy per site $E_N/N$ has a kink, signaling the level crossing in the system, see Fig.\ref{fig: first order Haldane and plaquette}(a).
This kink becomes even more pronounced if we consider the bulk energy per site $(E_{N_1}-E_{N_2})/(N_1-N_2)$ extracted from a finite-size results for two different system sizes, allowing us to effectively reduce boundary effects.
And second, we report a clear jump in the order parameter, shown in Fig.\ref{fig: first order Haldane and plaquette}(b), that shows no finite-size effects. 

\section{First order transition between Haldane and dimerized phases}\label{Appendix: first order Haldane and dimerized}
\begin{figure}[!h]
    \includegraphics{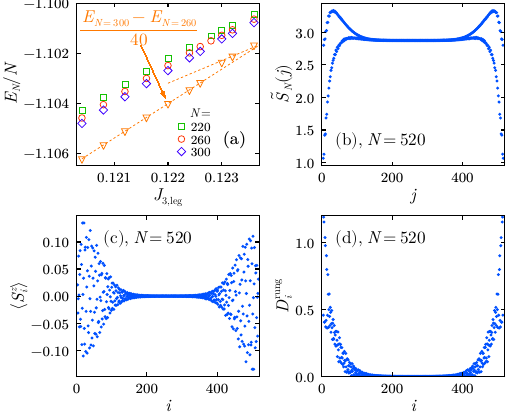}
    \caption{
        Data supporting a first order transition between the topologically non-trivial Haldane phase and the topologically trivial $\mathbb{Z}_2$ ordered dimerized phase at $J_1=0.5$.
        (a) Energy per site $E_N$ across the transition.
        The bulk energy per site, that we calculate as $(E_{N_1}-E_{N_2})/(N_1-N_2)$ (orange circles), shows two different slopes.         Two dashed line correspond to a fit to emphasize this.
 The discontinuity is due to a finite-size shift in the location of the transition.
        (b)-(d) Coexistence of Haldane and dimerized domains.
        Reduced entanglement entropy (b), local magnetization (c), and the dimerization on the rungs indicating dimerized domain near the edges and a central domain in the Haldane phase.    }
    \label{fig: first order Haldane and dimerized}
\end{figure}

Signatures of a first order transition between the dimerized and Haldane phase are less pronounced than the transition between the Haldane and plaquette phase. 
The energy per site $E_N$, for example, that we show in Fig.\ref{fig: first order Haldane and dimerized}(a) does not have a clear kink, hinting at the presence of weak first order behavior. 
Nonetheless, the two distinct slopes observed in the bulk energy points towards a level crossing and a first order transition \footnote{The discontinuity in the bulk energy is a result of the finite-size effect shifting away the location of the transition.}. 
In addition, we observe a clear signature of the coexistence of Haldane and dimerized domains in Fig.\ref{fig: first order Haldane and dimerized}(b)-(d).
Near the edges of the chain we find domains that feature spontaneous dimerization, while in the central domain we find an absence of such dimerization, as is expected for the Haldane phase.
The location of where the dimerization starts to increase matches the peaks of the entanglement entropy associated with the location of domain walls. 
Similar domain coexistence has been previously reported in the context of quantum loop model \cite{Z4_transition_QLM}.

\section{Diverging correlation length at the Gaussian transition}\label{Appendix: Gaussian}

\begin{figure}[!h]
    \includegraphics{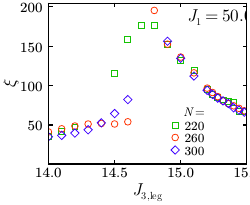}
    \caption{
        Diverging correlation length along a cut through the Gaussian transition at $J_1=50$.
        Correlation length is extracted from the spin-spin correlation function and shown for three chain lengths.
        Deviations from the diverging trend is due to challenges in fitting incommensurate oscillations.
    }
    \label{fig: Gaussian appendix}
\end{figure}

While the Gaussian transition can not be characterized through a local order parameter in this case, we can still recognize the critical phenomena that corresponds to the breaking of the topological order inherent to the Haldane phase through a diverging correlation length. 
In Fig.\ref{fig: Gaussian appendix} we show the correlation extracted from the spin-spin correlation function $C_{i,j} = \langle S_i^z S_j^z \rangle - \langle S_i^z \rangle \langle S_j^z \rangle$.
One can see a clear divergence of the correlation length.

\section{Extracting $\xi$ from the correlation function}\label{Appendix: correlation function}
\begin{figure}[!h]
    \includegraphics{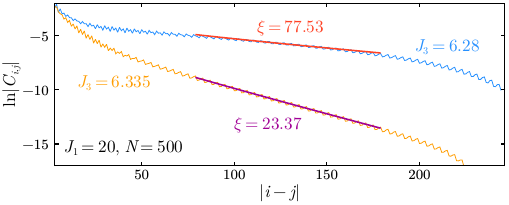}
    \caption{
        Example of fitting the correlation function $C_{i,j}$.
        From the slope of $\ln |C_{i,j}|$ we extract the correlation length $\xi$.
    }
    \label{fig: correlation function fit}
\end{figure}

To obtain the correlation length $\xi$, we fit the numerically computed connected correlation function, such as the leg-leg correlation function defined in Eq.\eqref{eq: connect correlations}, to Eq.\eqref{eq: correlation length}.
For this, we plot the logarithm of the correlation function, as shown in Fig.\ref{fig: correlation function fit}, and perform a linear fit of its slope with the function $\ln |C_{i,j}| = A - |i-j|/\xi - \ln(|i-j|)/2$, with $\xi$ and the non-universal amplitude $A$ as fitting parameters.
We always examine the fitting range carefully due to the possibility of incommensurate short range order (see Appendix \ref{Appendix: incommensurate oscillations} for an example).

\section{Incommensurate oscillations}\label{Appendix: incommensurate oscillations}
\begin{figure}[!h]
    \includegraphics{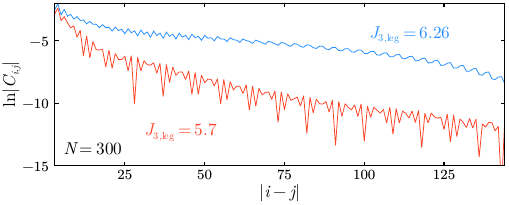}
    \caption{
        Leg-leg correlation function, defined in Eq.\eqref{eq: connect correlations}, for two values of the three-body interaction $J_{3,\mathrm{leg}}$ in a semi-log scale.
        We present data for $J_1=20$ and $N=300$ sites.
        Incommensurate oscillations are clearly present for $J_{3,\mathrm{leg}}=5.7$, while close to the Ashkin-Teller transition (data for $J_{3,\mathrm{leg}}=6.26$) the curves are nearly commensurate.
    }
    \label{fig: incommensurate oscillations}
\end{figure}

Incommensurate short range order is clearly present in the leg-leg correlation function shown in Fig.\ref{fig: incommensurate oscillations}.
These incommensurate oscillations are visible over a large range of $J_{3,\mathrm{leg}}$, and they become commensurate over a large crossover region that spans the width of the intermediate disorder phase.
In other words, the Ashkin-Teller transition studied here is truly a commensurate-commensurate transition.

\section{Additional data Ashkin-Teller transition}\label{Appendix: Ashkin-Teller}
\begin{figure*}
    \includegraphics{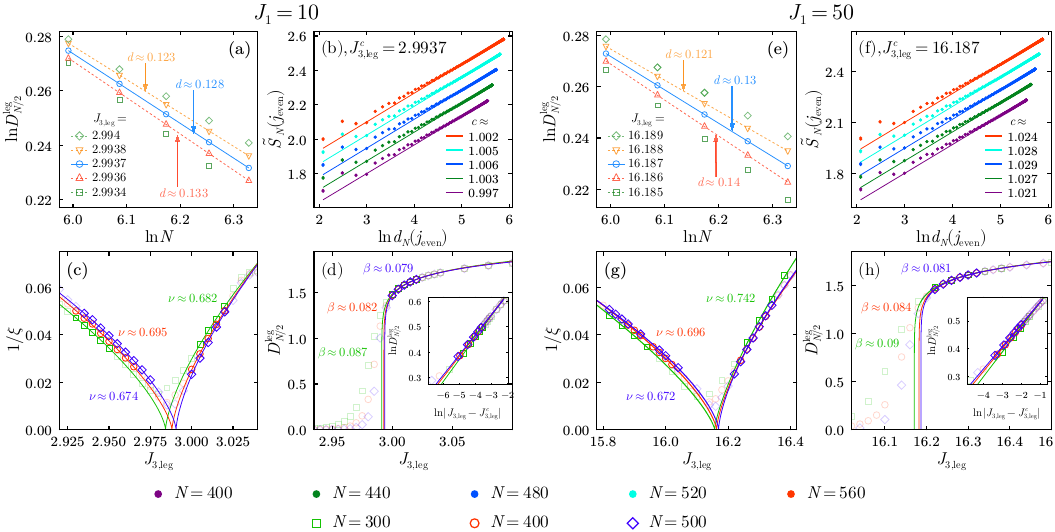}
    \caption{
        Additional data for the Ashkin-Teller transition line between the topological trivial disordered and $\mathbb{Z}_4$ ordered phases.
        Panels (a)-(d) shows data for $J_1=10$.
        (a) Finite size scaling of the dimerization on the legs, defined in \eqref{eq:dimerization_order_parameter}, in the middle of the chain in a log-log scale.
        We show data for five different values of $J_{3,\mathrm{leg}}$ (colored symbols).
        Lines indicate linear fits of the data.
        From the separatrix at $J_{3,\mathrm{leg}}^c = 2.9937$ -- denoting the critical point -- we extract a scaling dimension $d \approx 0.130$ that is in excellent agreement with the theoretical value $d=1/8$.
        Concave and convex curves are slightly shifted for visual clarity.
        (b) Reduced entanglement entropy as a function of the logarithm of the conformal distance at the critical point $J_{3,\mathrm{leg}}^c$ (colored dots).
        We only show data for bi-partitions made through the middle of a plaquette, or between two subsequent ones.
        Except for $N=400$, curves are shifted for visual clarity.
        We extract a central charge $c$ that is within a 3\% error margin of $c=1$ for the Ashkin-Teller class. 
        (c) Inverse of the correlation length for three system sizes (colored symbols).
        Lines depict $\xi \propto |J_{3,\mathrm{leg}}-J_{3,\mathrm{leg}}^c|^{-\nu}$ fitted to the data; $\nu$ lies within $2/3 \leq \nu \leq 1$ that characterizes the family of Ashkin-Teller models.
        Data with $\xi > N/10$ and that far from the critical point are excluded from the fit (greyed out data).
        (d) Mid-chain dimerization on the legs in vicinity to the critical point.
        Curves show a fit of the data to $D^\mathrm{leg}_{N/2} \propto (J_{3,\mathrm{leg}}-J_{3,\mathrm{leg}}^c)^{\beta}$.
        We extract a critical exponent $\beta$ matching $1/12 \leq \beta \leq 1/8$.
        Greyed out data is excluded from the fits.
        Inset: Same data but in a log-log scale.
        Fitted data appears as linear.
        (e)-(f) Same as in (a)-(d) but for $J_1=50$.
        Note, the critical exponents, their relation, and the central charge does not show in significant changes from the data shown for $J_1=10$ or $J_1=20$ (see Fig.\ref{fig: Ashkin-Teller}).
    }
    \label{fig: Ashkin-Teller appendix}
\end{figure*}

In Fig.\ref{fig: Ashkin-Teller appendix} we present data supporting Ashkin-Teller criticality between the topologically trivial disordered phase and the topologically trivial $\mathbb{Z}_4$ ordered plaquette phase for two cuts at $J_1=10$ and $J_1=50$.
Rather surprisingly, in all presented cases the non-universal critical exponents $\nu$ and $\beta$  do not show any noticeable dependence  on the value of $J_1$.

\clearpage

\bibliography{Paper}


\clearpage

\end{document}